\documentclass[
aps,%
12pt,%
final,%
notitlepage,%
oneside,%
onecolumn,%
nobibnotes,%
nofootinbib,% 
superscriptaddress,%
noshowpacs,%
centertags]%
{revtex4-2}
\providecommand{\refitem}[1]{}
\usepackage{indentfirst}
\usepackage{hyperref}
\usepackage{color}
\usepackage{units}
\usepackage{amssymb}
\usepackage{amsmath}
\usepackage[english]{babel}
\usepackage{graphicx}
\usepackage{float}
\usepackage{braket}
\usepackage{tikz-feynman}
\usepackage{feynmp-auto} %диаграммы Феймана 
\usepackage{tikz}
\usetikzlibrary{arrows.meta}
\usetikzlibrary{decorations.pathmorphing}
\usetikzlibrary{shapes.geometric}
\usetikzlibrary{decorations.text}
\tikzset{snake it/.style={decorate, decoration={snake, segment length=3mm, amplitude=0.4mm}}}
\tikzset{oplus/.style={path picture={
			\draw[black, line width=1pt]
			(path picture bounding box.south west) -- (path picture bounding box.north east)
			(path picture bounding box.north west) -- (path picture bounding box.south east);
}}}
\tikzset{ooplus/.style={path picture={
			\node at (path picture bounding box.center)
			[circle, color=white, fill=black]{};
			\draw[white, line width=1.5pt]
			(path picture bounding box.south west) -- (path picture bounding box.north east)
			(path picture bounding box.north west) -- (path picture bounding box.south east);
}}}

\newcommand{\lam}{\lambda}
\newcommand{\vf}{\varphi}
\newcommand{\f}{\phi}

\begin{document}

% \selectlanguage{english} % Для статьи на русском языке
% \selectlanguage{english} % Для статьи на английском языке
%
\title{Mass quench in interacting finite-size quantum field system}
\author{G.~A.~Simonian}
\email{simonyan.grachik@gmail.com}
\affiliation{Lebedev Physical Institute, Russian Academy of Sciences, Moscow, Russia}
\author{A.~G.~Mikhaylenko}
\email{mikhajlenko.ag@phystech.edu}
\affiliation{Lebedev Physical Institute, Russian Academy of Sciences, Moscow, Russia}
\author{A.~G.~Semenov}
\email{semenovag@lebedev.ru}
\affiliation{Lebedev Physical Institute, Russian Academy of Sciences, Moscow, Russia}
\noaffiliation
\begin{abstract}
In this work, we theoretically investigate the non-equilibrium evolution of the correlation functions of a quantum scalar field following a mass quantum quench in a finite-size system. The self-consistent equation for the Keldysh (statistical) correlation function is solved numerically in the one-loop approximation. A specific spatio-temporal effect is observed: at a point in time approximately equal to half the system size, the effective mass experiences a strong disturbance. It is shown that this effect is driven by the interference of counter-propagating wave fronts of fluctuations that propagate within the light cone and is induced by periodic boundary conditions.
\end{abstract}
\maketitle
\section{introduction}
The study of highly nonequilibrium behavior of isolated quantum macrosystems is one of the most dynamically developing areas of modern condensed matter physics and quantum field theory (QFT) \cite{Berges:2015kfa, kamenev2023field, Arseev2015, batini2024real}. One of the most thoroughly studied methods of bringing a system into a non-equilibrium state is a quantum quench, an instantaneous, step-like change in one of the Hamiltonian's parameters (for instance, the coupling constant or mass) \cite{calabrese2007quantum, mitra2018quantum, radovskaya2023local, PismaZhETF.124.152, ageev2024local, belkovich2026global}. The Keldysh–Schwinger formalism is the theoretical foundation for describing post-quench evolution~\cite{Akhmedov,Keldysh1964,sieberer2016keldysh, batini2024real}.

An adequate description of such systems requires a formalism capable of simultaneously and self-consistently accounting for both the change in the energy spectrum of elementary excitations (spectral characteristics) and the evolution of the population of these modes (statistical characteristics). To address this problem, the present work considers the Dyson equations obtained within the Keldysh–Schwinger technique in a self-consistent one-loop approximation.

Quantum quench has previously been considered in such systems as conformal field theory~\cite{das2014universal}, the Hubbard model \cite{schiro2010time}, the Ising model \cite{calabrese2007quantum}, as well as in other spin \cite{PhysRevB.98.205145}, bosonic \cite{sotiriadis2010quantum} and fermionic theories \cite{liu2019complexity, roux2009quenches, li2023quantum}. A distinctive feature of this paper is that we consider a finite-size system and thoroughly investigate the effects arising from periodic boundary conditions.

In Section \ref{Sec:2}, we define the model under consideration and derive the equations for the correlation functions. In Section \ref{Sec:3}, we formulate the initial conditions and set the values of the numerical parameters. In Section \ref{Sec:4}, we present the main results related to the effect of periodic boundary conditions and explain the effect using the spectral function in coordinate representation. In Section \ref{Sec:5}, to verify the numerical computation scheme, we provide a derivation of an alternative formula for the Keldysh correlation function that includes the full spectral function. In Section \ref{Sec:6}, we discuss the results obtained.

\section{THEORETICAL FORMALISM AND KELDYSH TECHNIQUE}\label{Sec:2}
We consider a one-dimensional real scalar quantum field theory in a system of length $L$ with periodic boundary conditions described by the Hamiltonian:
\begin{equation}
    \hat{H} = \frac{1}{2}\int \mathrm{d} x \left[\hat{\pi}^2+(\partial_x\hat{\vf})^2+ m^2(t)\hat{\vf}^2\right] + \frac{\lam}{4!}\int \mathrm{d}x\,\hat{\vf}^4.
    \label{eq:Hamiltonian}
\end{equation}
Here $\hat{\vf}(t,x)$ and $\hat{\pi}(t,x)$ are the field and canonically conjugate momentum operators, satisfying the canonical commutation relations; $\lam>0$ is the self-coupling constant. The time dependence of the mass is given as a step function modeling the quench:
\begin{equation}
    m(t) = 
    \begin{cases} 
    m_0, & t \le 0, \\ 
    m_1, & t > 0. 
    \end{cases}
\end{equation}
Below, we present the derivation of the equations for the correlation functions, analogous to that performed in \cite{MSThMJ}. The evolution of the expectation value of the observable $\hat{O}=O(\hat{\varphi})$ is given by the expression:
\begin{equation}
	\mathrm{tr}\{\hat{\rho}_0\hat{O}^H(t)\} = \mathrm{tr}\{ \hat{\rho}_0\hat{U}^\dagger(t,0)\hat{O}\hat{U}(t,0) \}, \label{trO}
\end{equation}
where $\hat{\rho}_0$ is the density matrix operator at the initial moment of time $t=0$, and $\hat{U}(t,0)$ is the time-evolution operator. The matrix elements of the evolution operators from $0$ to $t$ and back can be expressed as path integrals over the fields $\varphi^F(t,x)$ and $\varphi^B(t,x)$, where the indices $F$ and $B$ denote the forward and backward branches of the closed time path (the Keldysh contour): 
\begin{figure}[H]
		\centering
		\begin{tikzpicture}[out=0,in=0,distance=0.25cm]
			\node (rho_0) at (0,0) {$\hat{\rho}_0$};
			\node[circle,fill,inner sep=1pt,scale=0.2,label=right:$\hat{O}$](O) at (9.5,0.04) {$\hat{O}$};
			\draw[arrows = {-Latex[round][line width=1pt, fill=black, length=5pt]}] (0.4,+0.15) -- node[above] {$\hat{U}(t,0)$} (5,+0.15);
			\draw (5,+0.15) -- node[above] {forward} (9.3,+0.15);
			\draw (9.3,+0.15) to (9.3,-0.15);
			\draw[arrows = {-Latex[round][line width=1pt, fill=black, length=5pt]}] (9.3,-0.15) -- node[below] {backward} (5,-0.15);
			\draw (5,-0.15) -- node[below] {$\hat{U}^\dagger(t,0)$} (0.4,-0.15);
		\end{tikzpicture}
\end{figure}
For convenience, we introduce symmetrized ("classical") $\varphi^{cl}$ and antisymmetrized ("quantum") $\varphi^q$ integration variables in the spacetime functional integral (hereafter $\hbar=1$):
\begin{equation}
    \vf^{cl}(t,x) = \frac{\vf^F(t,x)+\vf^B(t,x)}{2}, \quad \vf^q(t,x) = \vf^F(t,x)-\vf^B(t,x).
\end{equation}
In terms of these variables, the expectation value of the observable can be expressed as a path integral over spacetime field configurations:
\begin{equation}
\operatorname{tr}\left\{\hat{\rho}_{0} \hat{O}^{H}(t)\right\}=\int D \varphi^{c l} D \varphi^{q} \  W\left(\vf^{cl}(0),\dot{\vf}^{cl}(0)\right) \, O\left(\varphi^{c l}(t)\right) \exp \left\{i S^{K}\right\}.  
\end{equation}
Here, the initial state Wigner functional, described by the density matrix $\hat{\rho}_0$, is given by:
\begin{equation}
W(\varphi,\pi) = \int \mathcal{D}\phi~\Bra{\varphi+\frac{\phi}{2}}\hat{\rho}_0\Ket{\varphi-\frac{\phi}{2}}~\exp\left\{-i\int d^dx~\pi\phi\right\},
\end{equation}
where the integration is performed over spatial configurations of the real scalar field $\phi(x)$, which is an eigenstate of the operator $\hat{\phi}(x)$: $\hat{\phi}(x) \ket{\f} = \f(x)\ket{\f}$.
The Keldysh action, defined as the difference of classical actions on the two branches of the closed time path, in this case takes the following form:
\begin{equation}
    S^{K} = -\int_{0}^{\infty} \mathrm{d} t\int \mathrm{d}x\,\vf^{q}\left( \partial_t^2 - \partial_x^2 + m_1^2 \right)\vf^{c l} - \frac{\lambda}{6} \int_{0}^{\infty} \mathrm{d} t \int \mathrm{d}x\, \vf^q {(\vf^{cl})}^3 - \frac{\lambda}{24}\int_{0}^{\infty} \mathrm{d} t \int \mathrm{d}x \,(\vf^q)^3 \vf^{cl}.
\end{equation}

The main objects of research are two-point correlation functions. The Keldysh (statistical) Green's function is given by the expression:
\begin{equation}
    \langle \vf^{cl}(t_1, x_1) \vf^{cl}(t_2, x_2) \rangle =: F(t_1, x_1; t_2, x_2) = \frac{1}{2} \mathrm{tr}\left\{ \hat{\rho}_0 \{ \hat{\vf}(t_1, x_1), \hat{\vf}(t_2, x_2) \} \right\}.
\end{equation}
The Green's spectral function $\rho(t_1, x_1;t_2, x_2)$, which determines the causal dynamics and commutation relations, is expressed as:
\begin{equation}
    \langle \vf^{cl}(t_1, x_1) \vf^{q}(t_2, x_2) \rangle =: -i  \theta(t_1-t_2)\rho(t_1, x_1; t_2, x_2) =  \theta(t_1-t_2)\mathrm{tr}\left\{ \hat{\rho}_0 [ \hat{\vf}(t_1, x_1), \hat{\vf}(t_2, x_2) ] \right\},
\end{equation}
where $\theta(t_1-t_2)$ is the Heaviside function. The Keldysh and spectral functions possess the following (anti)symmetry properties:
\begin{align}
	& F(t_1,x_1;t_2,x_2) = F(t_2,x_2;t_1,x_1), \notag\\
	&\rho(t_1,x_1;t_2,x_2) = -\rho(t_2,x_2;t_1,x_1).
\end{align}

For a spatially uniform distribution with a zero mean field ($\langle\vf\rangle=0$), applying the two-particle-irreducible effective action formalism or, equivalently, the partial resummation procedure of diagrams entering the Dyson–Schwinger equations in the one-loop (Hartree–Fock) approximation leads to a closed system of differential equations. The explicit form of the Dyson equations in this approximation is:
    \begin{equation}
		\begin{tikzpicture}[baseline={([yshift=-.5ex]current bounding box.center)}]
			\draw[double] (0.75,0)--(0,0);
		\end{tikzpicture}
		=~
		\begin{tikzpicture}[baseline={([yshift=-.5ex]current bounding box.center)}]
			\draw (0.75,0)--(0,0);
		\end{tikzpicture}
		+
		\begin{tikzpicture}[baseline={([yshift=-.5ex]current bounding box.center)}]
			\node[draw, circle, double, thick, fill=black!10] (sigma_r) at (1.25,0) {$\bold{\Sigma}^R$};
			\draw[arrows = {-Latex[round][line width=1pt, fill=black, length=5pt]}] (sigma_r) -- (0.25,0);
			\draw (0.25,0)--(0,0);
			\draw[double] (2.5,0)--(sigma_r);
		\end{tikzpicture}
		+
		\begin{tikzpicture}[baseline={([yshift=-.5ex]current bounding box.center)}]
			\node[draw, circle, double, thick, fill=black!10] (sigma_a) at (1.25,0) {$\bold{\Sigma}^A$};
			\draw[double, arrows = {-Latex[round][line width=1pt, fill=black, length=5pt]}] (sigma_a) -- (2.25,0);
			\draw[double] (2.25,0)--(2.5,0);
			\draw (0,0)--(sigma_a);
		\end{tikzpicture}
		\label{Dyson_2}
	\end{equation}
    \begin{equation}
		\begin{tikzpicture}[baseline={([yshift=-.5ex]current bounding box.center)}]
			\draw[double, arrows = {-Latex[round][line width=1pt, fill=black, length=5pt]}] (0.75,0) -- (0.25,0);
			\draw[double] (0.25,0)--(0,0);
		\end{tikzpicture}
		=
		\begin{tikzpicture}[baseline={([yshift=-.5ex]current bounding box.center)}]
			\draw[arrows = {-Latex[round][line width=1pt, fill=black, length=5pt]}] (0.75,0) -- (0.25,0);
			\draw (0.25,0)--(0,0);
		\end{tikzpicture}
		+
		\begin{tikzpicture}[baseline={([yshift=-.5ex]current bounding box.center)}]
			\node[draw, circle, double, fill=black!10, thick] (sigma_r) at (1.25,0) {$\bold{\Sigma}^R$};
			\draw[arrows = {-Latex[round][line width=1pt, fill=black, length=5pt]}] (sigma_r) -- (0.25,0);
			\draw (0.25,0)--(0,0);
			\draw[double,arrows = {-Latex[round][line width=1pt, fill=black, length=5pt]}] (2.5,0) -- (2,0);
			\draw[double] (2,0)--(sigma_r);
		\end{tikzpicture}
		\label{Dyson_1}
	\end{equation}
    \begin{equation}
		\begin{tikzpicture}[baseline={([yshift=-.5ex]current bounding box.center)}]
			\node[draw, circle,double,thick,fill=black!10] (sigma_r) at (0,0) {$\bold{\Sigma}^R$};
			\draw[densely dotted] (sigma_r) -- (-0.75,0);
			\draw[densely dotted] (sigma_r) -- (+0.75,0);
		\end{tikzpicture}
		~=~
		\begin{tikzpicture}[baseline={([yshift=-2ex]current bounding box.center)}]
			\draw[double] (2,0) to [bend left=45] (1.75,0.25);
			\draw[double] (1.75,0.25) to [bend left=45] (2,0.5);
			\draw[double] (2,0.5) to [bend left=45] (2.25,0.25);
			\draw[double] (2.25,0.25) to [bend left=45] (2,0);
			\filldraw (2,0) circle (1.5pt);
			\draw[densely dotted] (2,0) -- (1.75,0);
			\draw[densely dotted] (2,0) -- (2.25,0);
		\end{tikzpicture}
		~+~o(\lam)
		\label{Sigma_R_2_loop}
	\end{equation}
Here, a straight line is a Keldysh function $\langle \vf^{cl}(t_1, x_1) \vf^{cl}(t_2, x_2) \rangle$, and an arrow line is a retarded function $\langle \vf^{cl}(t_1, x_1) \vf^{q}(t_2, x_2) \rangle$. The thin line corresponds to the free function, and the double line to the interacting one. The advanced and retarded self-energies are related by $\Sigma^A(t_1, x_1; t_2, x_2) = \Sigma^R(t_2, x_2; t_1, x_1)$.

For numerical simulations, the spatial coordinate is discretized on a lattice containing $N$ sites with spacing $a$, so that the total size of the system is $L = a N$. In momentum space, the Dyson equations in the one-loop approximation for the spatially homogeneous case reduce to the following system of equations:
\begin{align}
    &\left( \partial_{t_1}^2 + M^2_{\text{eff}}(t_1) + \frac{4}{a^2} \sin^2{\frac{a p}{2}}\right) F_p(t_1, t_2) = 0,
    \label{eq:DysonF} \\ 
    &\left( \partial_{t_1}^2 + M^2_{\text{eff}}(t_1) + \frac{4}{a^2} \sin^2{\frac{a p}{2}} \right) \rho_p(t_1, t_2) = 0.
\end{align}

The effective mass $M_{\text{eff}}(t)$ is determined self-consistently via the sum over all momentum modes of the Keldysh correlation function at equal times:
\begin{equation}
    M_{\text{eff}}^2(t) = m_1^2 + \delta m^2 + \frac{\lambda}{2L}\sum_{q}F_{q}(t,t).
    \label{eq:Meff}
\end{equation}
Here, $\delta m^2 = - \frac{\lam}{4L} \sum_q \frac{1}{\omega_{1,q}}$ is the counterterm necessary to cancel the divergence arising in the last term. We show the origin of this divergence using the example of the free Keldysh correlation function calculated via the method of Bogoliubov coefficients. For the case of interest, where no interaction was present before the quench and the system was in thermal equilibrium at temperature $T$, this function is given by:
\begin{align}
    F_{0,p}(t_1, t_2) = \frac{\coth\left(\frac{\omega_{0,p}}{2T}\right)}{4\omega_{1,p}} \left[ \left( \frac{\omega_{1,p}}{\omega_{0,p}} + \frac{\omega_{0,p}}{\omega_{1,p}} \right) \cos\big(\omega_{1,p}(t_1 - t_2)\big) + \left( \frac{\omega_{1,p}}{\omega_{0,p}} - \frac{\omega_{0,p}}{\omega_{1,p}} \right) \cos\big(\omega_{1,p}(t_1 + t_2)\big) \right]. 
\end{align}
Here, $\omega_{0,p}$ and $\omega_{1,p}$ denote the frequencies of the mode with momentum $p$ before and after the quench (neglecting interaction), respectively:
\begin{equation}
    \omega_{0,p} = \sqrt{m_0^2 + \frac{4}{a^2}\sin^2\frac{pa}{2} },
\end{equation}
\begin{equation}
    \omega_{1,p} = \sqrt{m_1^2 + \frac{4}{a^2}\sin^2\frac{pa}{2} }.
\end{equation}
Since $F_{0,q} \propto \coth\left(\frac{\omega_{0,q}}{2T}\right) = 2n_{0,q} + 1$, where $n_{0,q}=1/(e^{\omega_{0,q}/T}-1)$, performing the momentum summation yields a divergent sum of vacuum zero-point energies. To cancel this divergence, the counterterm $\delta m^2$ mentioned above must be added. 

\section{INITIAL CONDITIONS AND NUMERICAL PARAMETERS}\label{Sec:3}

We consider the case where, before the quench ($t \le 0$), there was no interaction and the system was in a thermal state at temperature $T$ with a bare mass $m_0$.

Initial conditions for the Keldysh function $F_p(t_1, t_2)$ are determined by the Wigner functional of the initial state. In this case:
\begin{align}
    &F_p(0, 0) = \frac{1}{2\omega_{0,p}}\coth{\frac{\omega_{0,p}}{2T}},\label{eq:IC1} \\
    &\partial_{t_1} F_p(t_1, t_2) \big|_{t_1 = t_2 = 0} = 0,\label{eq:IC2} \\ 
    &\partial_{t_1} \partial_{t_2} F_p(t_1, t_2) \big|_{t_1 = t_2 = 0} = \frac{\omega_{0,p}}{2}\coth{\frac{\omega_{0,p}}{2T}}.\label{eq:IC3}
\end{align}

The initial conditions for the spectral function $\rho_p(t_1, t_2)$ are determined by the canonical commutation relations and do not depend on the initial state:
\begin{align}
    &\rho_p(0,0)=0, \label{eq:IC_rho1} \\
    &\partial_{t_1} \rho_p(t_1, t_2) \big|_{t_1 = t_2 = 0} = 1, \label{eq:IC_rho2} \\ 
    &\partial_{t_1}\partial_{t_2} \rho_p(t_1, t_2) \big|_{t_1 = t_2 = 0} =0. \label{eq:IC_rho3}
\end{align}

In the calculations, we employed the following dimensionless system parameters: number of lattice sites $N = 512$, coupling constant $\lam = 10$, pre-quench initial mass $m_0 = 1.5$, post-quench final mass $m_1 = 1$, and temperature $T=0.1$. To investigate effects related to periodic boundary conditions, different values of the spatial system size $L=10, 30, 60$ were analyzed.

\section{Interference effect and LIGHT CONE}\label{Sec:4}
\begin{figure}
	\centering
	\includegraphics[width=0.7\linewidth]{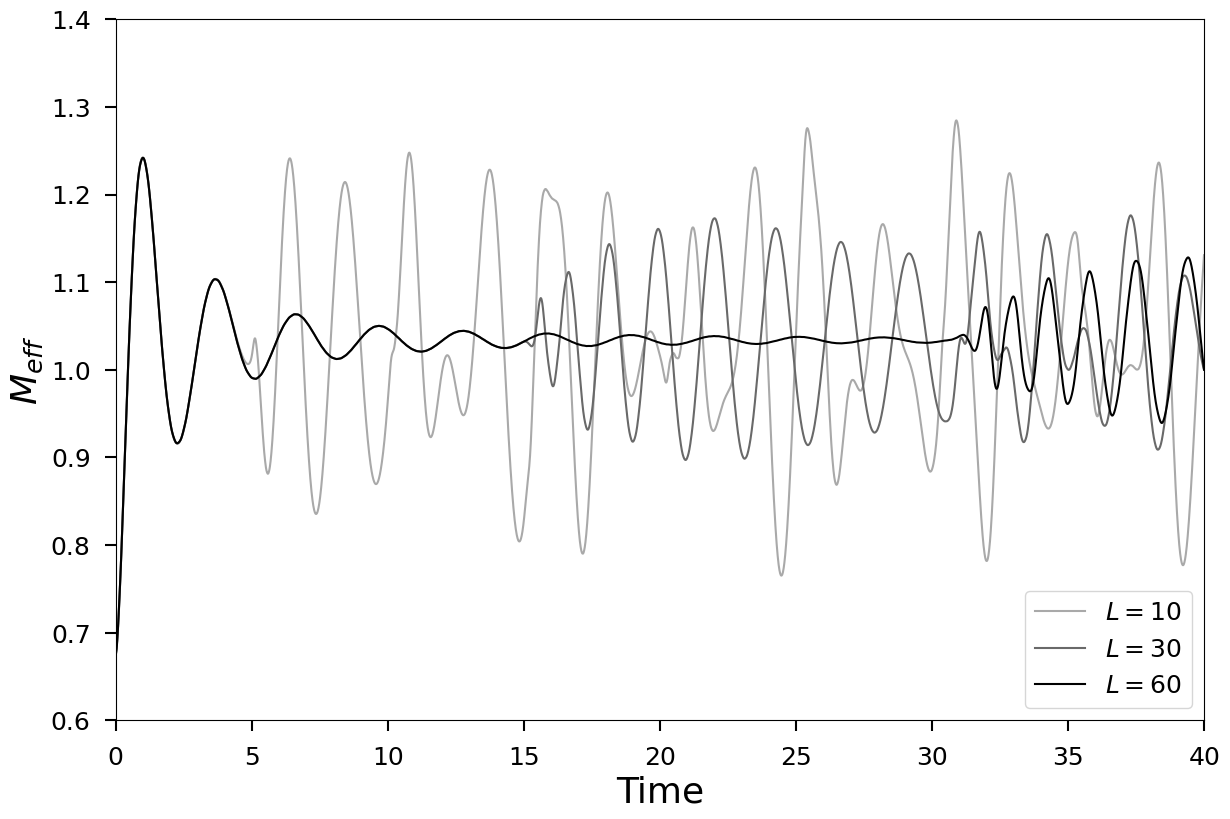}
	\caption{Time evolution of the effective mass  $M_{\text{eff}}(t)$ after a quantum quench depending on the system size $L = 10, 30, 60$.}
	\label{fig:quench1}
\end{figure}

The primary result of the numerical analysis of the system's post-quench evolution is the discovery of a sharp jump in the behavior of the effective mass $M_{\text{eff}}(t)$ at time $t \approx L/2$. The effective mass dynamics for different spatial ring sizes are shown in Fig.~\ref{fig:quench1}.
To effectively explain the physical nature of this phenomenon, it is useful to move from the momentum representation to a spatial analysis of the retarded Green's function on the ring.

Due to the periodic boundary conditions, the relation between the spectral function on the ring $\rho^{\text{ring}}_j(t) = \rho^{\text{ring}}(t,ja;0,0)$ (where $j$ is the spatial lattice site index) and the function on an infinite line $\rho^{\text{line}}_j(t)$ can be expressed as an infinite sum over sectors:
\begin{equation}
    \rho^{\text{ring}}_j(t)=\sum_{n=-\infty}^{\infty}\rho^{\text{line}}_{j+nN}(t).
    \label{eq:images}
\end{equation}

Due to microcausality in QFT, the spectral function on the line strictly vanishes outside the light cone (in units where $c=1$):
\begin{equation}
    \rho^{\text{line}}_{j+nN}(t) = 0 \quad \text{for} \quad |j + nN| \cdot a > t.
\end{equation}

\begin{figure}
\centering
\includegraphics[width=0.7\linewidth]{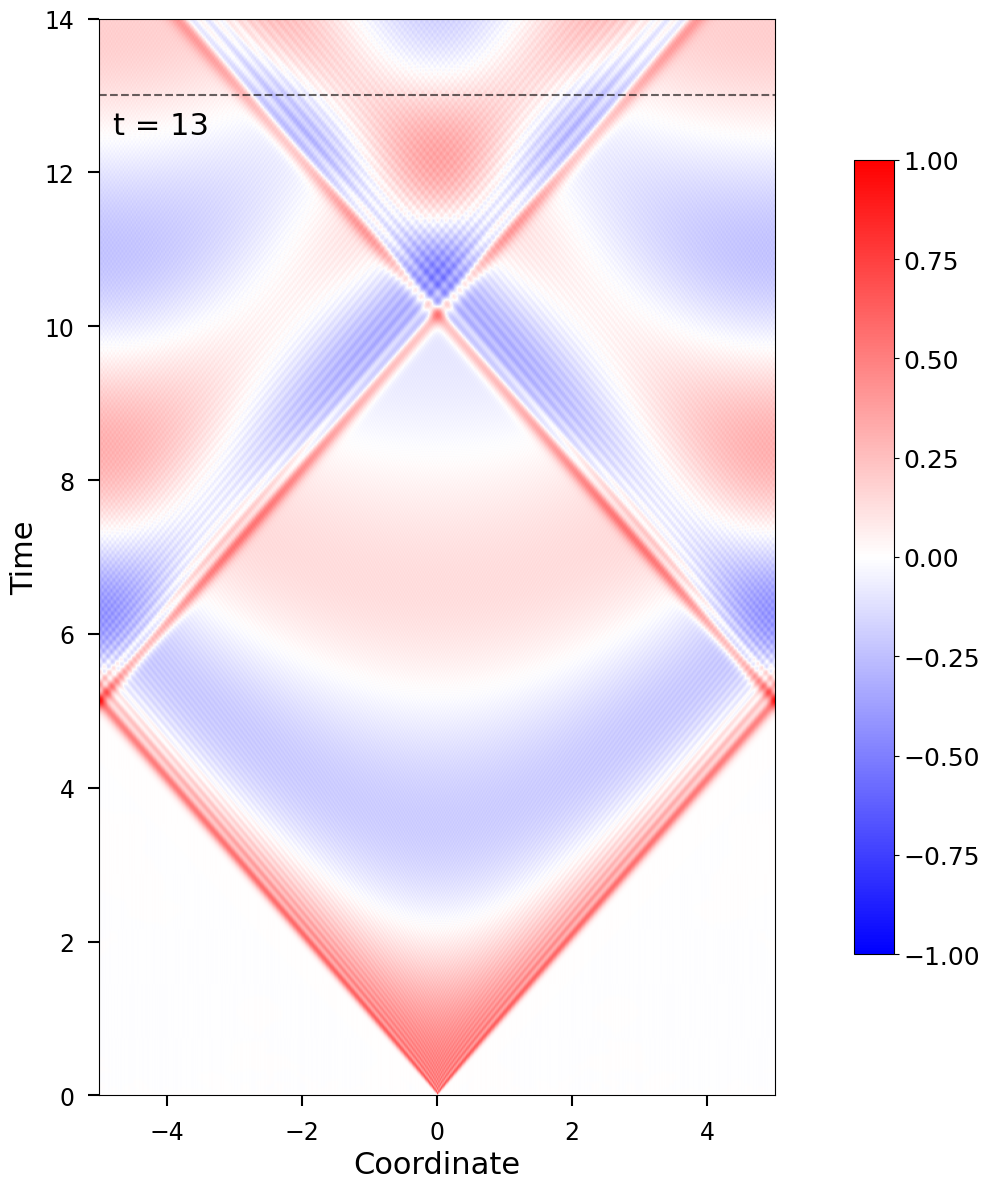}
\caption{Light cone of the retarded function}
\label{conus}
\end{figure}

As seen from Eq.~\eqref{eq:images}, at $t = L/2$, the summation index $n$ on the right-hand side takes not only the value zero, as was the case for $t < L/2$, but also $n = \pm 1$. The spacetime dependence of the retarded Green's function (exhibiting the characteristic light cone structure) is shown in Fig.~\ref{conus}.

\begin{figure}
		\centering
		\includegraphics[width=0.7\linewidth]{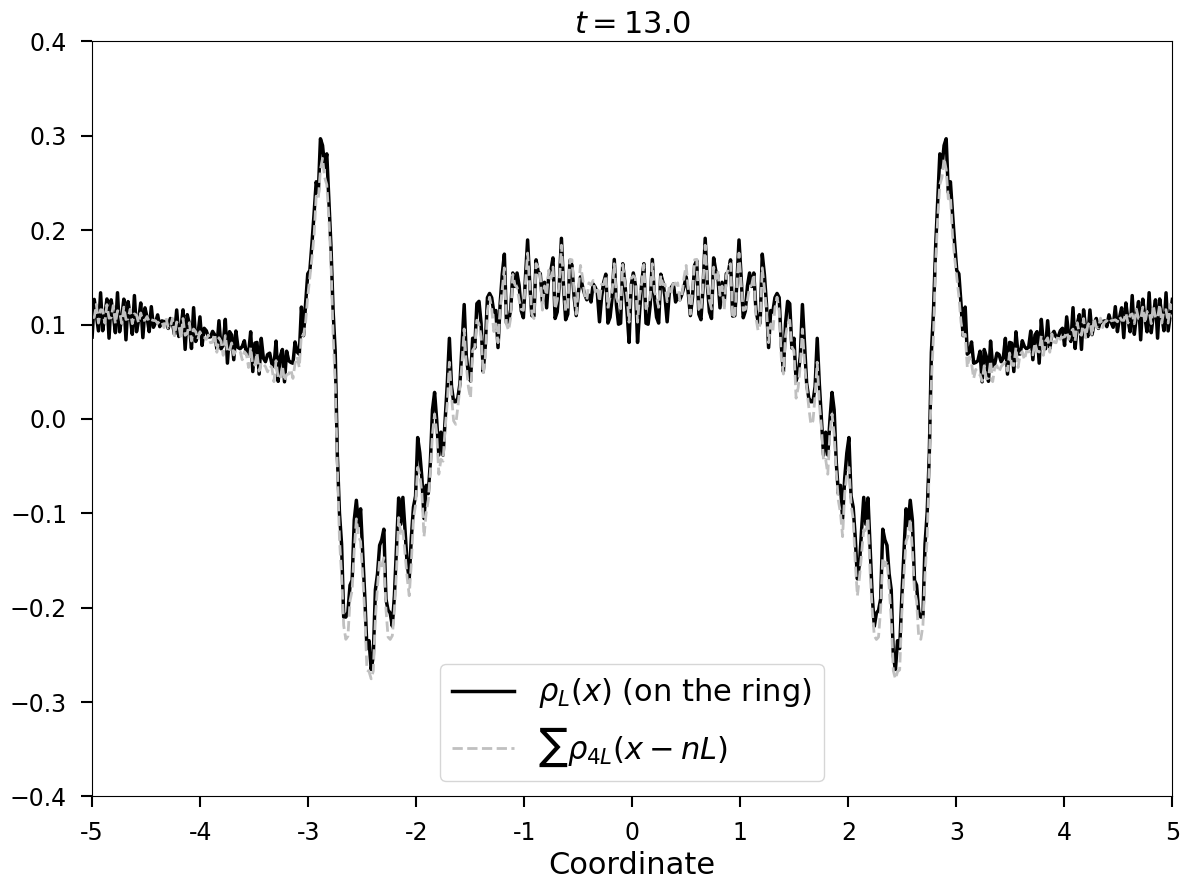}

	\caption{Light cone and comparison of retarded functions calculated via two different methods}
	\label{fig:Fpt}
\end{figure}

A comparison between the results obtained on the ring and on the infinite line is shown in Fig.~\ref{fig:Fpt}. The behavior of the retarded function on the line was simulated on a ring whose size was four times larger than the original one. As can be seen from the plots, the calculation results obtained by the two methods coincide with high precision. 

Thus, this effect is the result of interference between counter-propagating fluctuations that meet at the same point in space due to periodic boundary conditions.

\section{Alternative Computation of the Keldysh Function}\label{Sec:5}

To verify that the above-described change in the behavior of the effective mass is not an artifact of the numerical computation scheme, we calculate the Keldysh function using a different method. For this purpose, we obtain an expression from the Dyson equations for the spectral function in terms of the free Keldysh function and the interacting spectral function.

Let's assume that, having solved the Dyson equations numerically, we have already found a simultaneous Keldysh function $F_p(t,t)$. Then we can replace the loop in the self-energy with this computed value, which we denote by a square in the diagrams. Using the Dyson equation \eqref{Dyson_2}, in which the loop is replaced by a square, we obtain the identity:
\begin{equation}
    \begin{tikzpicture}[baseline={([yshift=-1.6ex]current bounding box.center)}]
		
        \draw[double] (0.75,0)--(0,0);
		\draw[double,arrows = {-Latex[round][line width=1pt, fill=black, length=5pt]}] (1,0) -- (1.25,0);
		\draw[double] (1.25,0)--(1.5,0);
		\draw[double] (1,0)--(0.75,0);	
		\node [oplus,draw=black,double, scale=0.7] (y) at (+0.75,+0.4) {};
		\path [draw=black,double,snake it] (y) -- (0.75,0);
		\filldraw (0.75,0) circle (1.5pt);	
	\end{tikzpicture}
    =~
    \begin{tikzpicture}[baseline={([yshift=-1.6ex]current bounding box.center)}]
		
		\draw (0.75,0) -- (0.25,0);
		\draw (0.25,0)--(0,0);
		\draw[double,arrows = {-Latex[round][line width=1pt, fill=black, length=5pt]}] (1,0) -- (1.25,0);
		\draw[double] (1.25,0)--(1.5,0);
		\draw[double] (1,0)--(0.75,0);	
		\node [oplus,draw=black,double, scale=0.7] (y) at (+0.75,+0.4) {};
		\path [draw=black,double,snake it] (y) -- (0.75,0);
		\filldraw (0.75,0) circle (1.5pt);	
	\end{tikzpicture}
    ~+~
    \begin{tikzpicture}[baseline={([yshift=-1.6ex]current bounding box.center)}]
		\draw[double, arrows = {-Latex[round][line width=1pt, fill=black, length=5pt]}] (0.75,0) -- (0.25,0);
		\draw[double] (0.25,0)--(0,0);
		\draw (1.5,0) -- (0.75,0);	
		\node [oplus,draw=black,double, scale=0.7] (y) at (+0.75,+0.4) {};
		\path [draw=black,double,snake it] (y) -- (0.75,0);
		\filldraw (0.75,0) circle (1.5pt);
		\draw[double,arrows = {-Latex[round][line width=1pt, fill=black, length=5pt]}] (1.5,0) -- (1.75,0);
		\draw[double] (1.75,0)--(2.,0);
		\draw[double] (1.5,0)--(1.25,0);	
		\node [oplus,draw=black,double, scale=0.7] (z) at (+1.25,+0.4) {};
		\path [draw=black,double,snake it] (z) -- (1.25,0);
		\filldraw (1.25,0) circle (1.5pt);
	\end{tikzpicture}
    ~+~
    \begin{tikzpicture}[baseline={([yshift=-1.6ex]current bounding box.center)}]
		\draw[double] (0.75,0) -- (0.25,0);
		\draw[double] (0.25,0)--(0,0);
		\draw[arrows = {-Latex[round][line width=1pt, fill=black, length=5pt]}] (0.75,0) -- (1.125,0);
        \draw (1.125,0) -- (1.25,0);
		\node [oplus,draw=black,double, scale=0.7] (y) at (+0.75,+0.4) {};
		\path [draw=black,double,snake it] (y) -- (0.75,0);
		\filldraw (0.75,0) circle (1.5pt);
		\draw[double,arrows = {-Latex[round][line width=1pt, fill=black, length=5pt]}] (1.5,0) -- (1.75,0);
		\draw[double] (1.75,0)--(2.,0);
		\draw[double] (1.5,0)--(1.25,0);	
		\node [oplus,draw=black,double, scale=0.7] (z) at (+1.25,+0.4) {};
		\path [draw=black,double,snake it] (z) -- (1.25,0);
		\filldraw (1.25,0) circle (1.5pt);
	\end{tikzpicture}
\end{equation}
Here we have rewritten the full correlation function on the left-hand side of the diagram using the Dyson equation \eqref{Dyson_2} mentioned above. Expressing the last diagram on the right-hand side using another Dyson equation \eqref{Dyson_1}, we obtain:
\begin{equation}
    \begin{tikzpicture}[baseline={([yshift=-1.6ex]current bounding box.center)}]
		
        \draw[double] (0.75,0)--(0,0);
		\draw[double,arrows = {-Latex[round][line width=1pt, fill=black, length=5pt]}] (1,0) -- (1.25,0);
		\draw[double] (1.25,0)--(1.5,0);
		\draw[double] (1,0)--(0.75,0);	
		\node [oplus,draw=black,double, scale=0.7] (y) at (+0.75,+0.4) {};
		\path [draw=black,double,snake it] (y) -- (0.75,0);
		\filldraw (0.75,0) circle (1.5pt);	
	\end{tikzpicture}
    =~
    \begin{tikzpicture}[baseline={([yshift=-1.6ex]current bounding box.center)}]
		
		\draw (0.75,0) -- (0.25,0);
		\draw (0.25,0)--(0,0);
		\draw[double,arrows = {-Latex[round][line width=1pt, fill=black, length=5pt]}] (1,0) -- (1.25,0);
		\draw[double] (1.25,0)--(1.5,0);
		\draw[double] (1,0)--(0.75,0);	
		\node [oplus,draw=black,double, scale=0.7] (y) at (+0.75,+0.4) {};
		\path [draw=black,double,snake it] (y) -- (0.75,0);
		\filldraw (0.75,0) circle (1.5pt);	
	\end{tikzpicture}
    ~+~
    \begin{tikzpicture}[baseline={([yshift=-1.6ex]current bounding box.center)}]
		\draw[double, arrows = {-Latex[round][line width=1pt, fill=black, length=5pt]}] (0.75,0) -- (0.25,0);
		\draw[double] (0.25,0)--(0,0);
		\draw (1.5,0) -- (0.75,0);	
		\node [oplus,draw=black,double, scale=0.7] (y) at (+0.75,+0.4) {};
		\path [draw=black,double,snake it] (y) -- (0.75,0);
		\filldraw (0.75,0) circle (1.5pt);
		\draw[double,arrows = {-Latex[round][line width=1pt, fill=black, length=5pt]}] (1.5,0) -- (1.75,0);
		\draw[double] (1.75,0)--(2.,0);
		\draw[double] (1.5,0)--(1.25,0);	
		\node [oplus,draw=black,double, scale=0.7] (z) at (+1.25,+0.4) {};
		\path [draw=black,double,snake it] (z) -- (1.25,0);
		\filldraw (1.25,0) circle (1.5pt);
	\end{tikzpicture}
    ~+~
    \begin{tikzpicture}[baseline={([yshift=-1.6ex]current bounding box.center)}]
		
        \draw[double] (0.75,0)--(0,0);
		\draw[double,arrows = {-Latex[round][line width=1pt, fill=black, length=5pt]}] (1,0) -- (1.25,0);
		\draw[double] (1.25,0)--(1.5,0);
		\draw[double] (1,0)--(0.75,0);	
		\node [oplus,draw=black,double, scale=0.7] (y) at (+0.75,+0.4) {};
		\path [draw=black,double,snake it] (y) -- (0.75,0);
		\filldraw (0.75,0) circle (1.5pt);	
	\end{tikzpicture}
    ~-~
    \begin{tikzpicture}[baseline={([yshift=-1.6ex]current bounding box.center)}]
		
        \draw[double] (0.75,0)--(0,0);
		\draw[arrows = {-Latex[round][line width=1pt, fill=black, length=5pt]}] (1,0) -- (1.25,0);
		\draw (1.25,0)--(1.5,0);
		\draw (1,0)--(0.75,0);	
		\node [oplus,draw=black,double, scale=0.7] (y) at (+0.75,+0.4) {};
		\path [draw=black,double,snake it] (y) -- (0.75,0);
		\filldraw (0.75,0) circle (1.5pt);	
	\end{tikzpicture}
\end{equation}
Expressing the last diagram from this and substituting it back into  Eq.~\eqref{Dyson_2}, we arrive at the desired expression for the Keldysh function $F_p(t_1,t_2)$: 
\begin{equation}
	\begin{tikzpicture}[baseline={([yshift=-.6ex]current bounding box.center)}]
		\draw[double] (0.75,0)--(0,0);
	\end{tikzpicture}
	=~
	\begin{tikzpicture}[baseline={([yshift=-.6ex]current bounding box.center)}]
		\draw (0.75,0)--(0,0);
	\end{tikzpicture}
	+
	\begin{tikzpicture}[baseline={([yshift=-1.6ex]current bounding box.center)}]
		\draw[double, arrows = {-Latex[round][line width=1pt, fill=black, length=5pt]}] (0.75,0) -- (0.25,0);
		\draw[double] (0.25,0)--(0,0);
		\draw (1.5,0) -- (1,0);
		\draw (1,0)--(0.75,0);	
		\node [oplus,draw=black,double, scale=0.7] (y) at (+0.75,+0.4) {};
		\path [draw=black,double,snake it] (y) -- (0.75,0);
		\filldraw (0.75,0) circle (1.5pt);
	\end{tikzpicture}
	+
		\begin{tikzpicture}[baseline={([yshift=-1.6ex]current bounding box.center)}]
		
		\draw (0.75,0) -- (0.25,0);
		\draw (0.25,0)--(0,0);
		\draw[double,arrows = {-Latex[round][line width=1pt, fill=black, length=5pt]}] (1,0) -- (1.25,0);
		\draw[double] (1.25,0)--(1.5,0);
		\draw[double] (1,0)--(0.75,0);	
		\node [oplus,draw=black,double, scale=0.7] (y) at (+0.75,+0.4) {};
		\path [draw=black,double,snake it] (y) -- (0.75,0);
		\filldraw (0.75,0) circle (1.5pt);	
	\end{tikzpicture}
	+
	\begin{tikzpicture}[baseline={([yshift=-1.6ex]current bounding box.center)}]
		\draw[double, arrows = {-Latex[round][line width=1pt, fill=black, length=5pt]}] (0.75,0) -- (0.25,0);
		\draw[double] (0.25,0)--(0,0);
		\draw (1.5,0) -- (0.75,0);	
		\node [oplus,draw=black,double, scale=0.7] (y) at (+0.75,+0.4) {};
		\path [draw=black,double,snake it] (y) -- (0.75,0);
		\filldraw (0.75,0) circle (1.5pt);
		\draw[double,arrows = {-Latex[round][line width=1pt, fill=black, length=5pt]}] (1.5,0) -- (1.75,0);
		\draw[double] (1.75,0)--(2.,0);
		\draw[double] (1.5,0)--(1.25,0);	
		\node [oplus,draw=black,double, scale=0.7] (z) at (+1.25,+0.4) {};
		\path [draw=black,double,snake it] (z) -- (1.25,0);
		\filldraw (1.25,0) circle (1.5pt);
	\end{tikzpicture}
	\label{eqF222}
\end{equation}
From this, we obtain an explicit expression for the equal-time Keldysh function:
\begin{align}
    &F_p(t,t) = F_{0,p}(t,t) - 2\int_0^t \mathrm{d}\tau \, \rho(t,\tau) \Sigma(\tau)F_{0,p}(\tau,t) \notag \\
    &\hspace{200pt}+\int_0^t \mathrm{d}\tau_1 \int_0^t \mathrm{d}\tau_2 \, \rho(t,\tau_1) \Sigma(\tau_1) F_{0,p}(\tau_1,\tau_2)\Sigma(\tau_2) \rho(\tau_2,t),
\end{align}
where $\Sigma(\tau) = M_{\text{eff}}^2(\tau) - m_1^2$. Comparing the two approaches shows that the solutions agree with very high precision, confirming that the observed behavior is not a numerical artifact.

\section{Discussion of Results}\label{Sec:6}
In the present paper, we investigated the non-equilibrium dynamics of the scalar $\vf^4$ quantum field model following a massive quantum quench in the spatially one-dimensional case. The key result of the numerical simulation is the discovery of a anomaly in the time dependence of the effective mass $M_{\text{eff}}(t)$ at time $t \approx L/2$.

As shown by the detailed theoretical analysis performed in Section 4, this effect is purely geometric in nature and is due to periodic boundary conditions. At time $t = L/2$, the counter-propagating wave fronts of quantum fluctuations, isotropically created at the moment of the quench, travel around exactly half of the ring and begin to interfere strongly with each other.

The conclusions obtained emphasize the importance of accounting for finite-size system effects when describing highly non-equilibrium phenomena, and can serve as a theoretical basis for interpreting precision experiments with ultracold atoms in periodic one-dimensional traps~\cite{Zenesini2024}, where finite-size effects are fundamentally unavoidable.

This work was supported by the Russian Science Foundation under grant No. 25-22-00832.

\bibliography{references}

\end{document}